\documentstyle[preprint,tighten,aps,epsfig]{revtex}

\begin{document}
\draft

%\twocolumn[\hsize\textwidth\columnwidth\hsize\csname@twocolumnfalse%
%\endcsname

\title {A nonextensive critical phenomenon scenario for quantum entanglement}

\author{Constantino Tsallis $^{(a,b)}$, Pedro W. Lamberti $^{(c)}$ and Domingo Prato $^{(c)}$}

\address{$^{(a)}$Centro Brasileiro de Pesquisas Fisicas, 
Xavier Sigaud 150, 22290-180, 
Rio de Janeiro-RJ, Brazil (tsallis@cbpf.br),\\
$^{(b)}$Erwin Schroedinger International Institute for Mathematical Physics, Boltzmanngasse 9, A-1090 Wien, Austria,\\
$^{(c)}$FaMAF, Universidad de Cordoba, Ciudad Universitaria, Cordoba, Argentina\\
%$\;\;\;\;\;\;\;\;\;\;\;\;\;\;\;\;\;\;\;\;\;\;\;\;\;\;\;\;\;\;\;\;\;\;\;\;\;\;\;\;\;\;\;\;\;\;\;\;\;\;\;\;\;\;\;\;\;\;\;\;\;\;\;\;\;\;\;\;\;\;\;\;\;\;\;\;\;\;\;\;\;\;\;\;\;\;\;\;\;\;\;\;\;\;\;\;\;\;\;\;$
$\;\;\;\;\;\;\;\;\;\;\;\;\;\;\;\;\;\;\;\;\;\;\;\;\;\;\;\;\;\;\;\;\;\;\;\;\;\;\;\;\;\;\;${\bf If a sailor does not know what harbor he is going to,\\
$\;\;\;\;\;\;\;\;\;\;\;\;\;\;\;\;\;\;\;\;\;\;\;\;\;\;\;\;\;\;\;\;\;\;\;\;\;\;\;\;\;\;\;\;\;\;\;\;\;$ for him no wind will be favorable}. [Lucius Annaeus Seneca]\\ 
}

\date{\today}

\maketitle

%\twocolumn

\begin{abstract}

We discuss the paradigmatic bipartite spin-1/2 system having the probabilities $\frac{1+3x}{4}$ of being in the Einstein-Podolsky-Rosen fully entangled state $|\Psi^-$$> \equiv \frac{1}{\sqrt 2}(|$$\uparrow>_A|$$\downarrow>_B$$-|$$\downarrow>_A|$$\uparrow>_B)$ and $\frac{3(1-x)}{4}$ of being orthogonal.
This system is known to be separable if and only if $x\le1/3$ (Peres criterion). This critical value has been recently recovered by Abe and Rajagopal through the use of the nonextensive entropic form $S_q \equiv \frac{1- Tr \rho^q}{q-1}\;(q \in \cal{R}; \; $$S_1$$= -$ $Tr$ $ \rho \ln \rho)$ which has enabled a current generalization of Boltzmann-Gibbs statistical mechanics. This result has been enrichened by Lloyd, Baranger and one of the present authors by proposing a critical-phenomenon-like scenario for quantum entanglement. Here we further illustrate and discuss this scenario through the calculation of some relevant quantities. 

\end{abstract}


\bigskip

\section{INTRODUCTION}

In the present paper we exhibit, for a very simple quantum system, namely two spins 1/2, a possible confluence of concepts coming from three distinct physical areas: nonextensive statistical mechanics, quantum entanglement and theory of critical phenomena. The formalism of nonextensive statistical mechanics \cite{tsallis} was proposed in 1988 by one of us, and is currently being applied \cite{levy,logistic,rafelski,beck1,ion,beck2} to a variety of thermodynamically anomalous systems which, in one way or another, exhibit (multi)fractals aspects. Among these anomalous systems, a prominent place is occupied by systems including long-range interactions. So being, it is after all not surprising that this thermostatistical formalism has interesting implications \cite{aberajagopal,others,sethmichel} in the area of quantum entanglement and its intrinsic nonlocality. Finally, quantum systems can be more or less entangled, which makes relevant the discussion of whether a given system is or not separable. Separability, which we shall define in detail later on, is a crucial feature in the discussion on whether a quantum physical system is susceptible of a {\it local realistic} description with hidden variables. These issues were first discussed in 1935 by Einstein, Podolsky and Rosen (EPR) \cite{EPR} and by Schroedinger \cite{schroedinger}, and since then by many others\cite{werner,GHZ,ekert,bennett,zurek2,quantumchaos,popescu,barenco,zurek,horodecki1,horodecki2,peres,horod,horodeckicube,popescu2,horodecki3,horodecki4,lloyd,bruss}. The very concept of {\it degree} of entanglement makes one naturally to think of the concept of order parameter in the theory of critical phenomena. Consequently, the above mentioned triple confluence emerges quite naturally.

Quantum entanglement is a quite amazing physical phenomenon, and has attracted intensive interest in recent years due to its possible applications in quantum computation, teleportation and cryptography, as well as to its connections to quantum chaos. As we shall see, entropic nonextensivity provides a path through which a critical phenomenon scenario can be established to quantitatively characterize quantum entanglement. In Section II we present, along Abe's lines\cite{abeaxiom}, the {\it classical} concept of nonextensive conditional entropy. In Section III we formulate its {\it quantum} version and use it, along the lines of \cite{aberajagopal,sethmichel}, to discuss a simple bipartite spin system. The results thus obtained enable the numerical calculation of relevant quantities which, exploring the degree of nonextensivity reflected by the entropic index $q$, make a critical phenomenon scenario to emerge for quantum entanglement . We finally conclude in Section IV.

\section{CLASSICAL CONDITIONAL nonextensive ENTROPY, $q$-EXPECTATION VALUES AND ESCORT DISTRIBUTIONS}

The classical version of nonextensive statistical mechanics is based on the following entropic form\cite{tsallis}:
\begin{equation}
S_q(\{p_i\}) \equiv \frac{1-\sum_{i=1}^W p_i^q}{q-1}\;\;\;(q\in {\cal R};\; W \ge 1;\;0 \le p_i \le1\;\forall i;\;\sum_{i=1}^W p_i=1)\;,
\end{equation}
which is nonnegative ($\forall q$), concave (convex) for $q>0$ ($q<0$), achieves the extremal value $\frac{W^{1-q}-1}{1-q}$ at equiprobability (a maximum if $q>0$ and a minimum if $q<0$), and reproduces, in the $q \rightarrow 1$ limit, the Boltzmann-Gibbs entropy $S_1 = - \sum_{i=1}^W p_i\;\ln p_i$. If $q<0$, the sum runs only on {\it strictly} positive probabilities $\{p_i\}$. 

Let us divide the set of $W$ possibilities in $U$ nonintersecting subsets respectively containing $W_1,\;W_2,\;...,\;W_U$ elements, with $\sum_{k=1}^U W_k = W$ ($U \le W$). We define the probabilities $\pi_1 \equiv \sum_{\{W_1\;terms\}} p_i$ , $\pi_2 \equiv \sum_{\{W_2\;terms\}} p_i$ , ..., $\pi_U \equiv \sum_{\{W_U\;terms\}} p_i$ , hence $\sum_{k=1}^U \pi_k = 1$. It is straightforward to verify the following generalization of the celebrated Shannon property:
\begin{equation}
S_q(\{p_i\}) = S_q(\{\pi_k\}) + \sum_{k=1}^U \pi_k^q S_q(\{p_i/\pi_k\})\;,
\end{equation}
where, consistently with Bayes' formula, $\{p_i/\pi_k\}$ are the {\it conditional} probabilities, and satisfy $\sum_{\{W_k\; terms\}} (p_i/\pi_k) = 1$. The nonnegative entropies $S_q(\{p_i\})$, $S_q(\{\pi_k\})$ and $S_q(\{p_i/\pi_k\})$ depend respectively on $W$, $U$ and $W_k$ probabilities. Eq. (2) can be rewritten as
\begin{equation}
S_q(\{p_i\}) = S_q(\{\pi_k\}) + \langle S_q(\{p_i/\pi_k\}) \rangle_q\;,
\end{equation}
where the {\it unnormalized $q$-expectation value} of the conditional entropy is defined as
\begin{equation}
\langle S_q(\{p_i/\pi_k\}) \rangle_q\ \equiv   \sum_{k=1}^U \pi_k^q S_q(\{p_i/\pi_k\})\;,
\end{equation}
Also, since the definition of $S_q(\{\pi_k\})$ implies
\begin{equation}
\frac{1+(1-q) S_q(\{\pi_k\})}{\sum_{k^{\prime}=1}^U \pi_{k^{\prime}}^q}=1 \;,
\end{equation}
Eq. (2) can be rewritten as follows:
\begin{equation}
S_q(\{p_i\}) = S_q(\{\pi_k\}) + \sum_{k=1}^U \pi_k^q \;  \frac{1+(1-q) S_q(\{\pi_k\})}{\sum_{k^{\prime}=1}^U \pi_{k^{\prime}}^q}\;          S_q(\{p_i/\pi_k\})\;.
\end{equation}
Consequently
\begin{equation}
S_q(\{p_i\}) = S_q(\{\pi_k\}) + \langle \langle S_q(\{\ p_i / \pi_k\})  \rangle \rangle_q +(1-q) S_q(\{\pi_k\})\langle \langle S_q(\{p_i / \pi_k\})  \rangle \rangle_q\;,
\end{equation}
where the {\it normalized $q$-expectation value} of the conditional entropy is defined as
\begin{equation}
\langle \langle S_q(\{ p_i / \pi_k \})  \rangle \rangle_q  \equiv  \sum_{k=1}^U  \Pi_k \;  S_q(\{ p_i / \pi_k \}) \;,
\end{equation}
with the {\it escort probabilities}\cite{beckescort} 
\begin{equation}
\Pi_k \equiv  \frac{\pi_k^q}{\sum_{k^{\prime}=1}^U \pi_{k^{\prime}}^q} \;\;\;(k=1,\;2,\;...,\;U) \;.
\end{equation}
Property (7) is, as we shall see later on, a very useful one, and it exhibits a most important fact, namely that the definition of the nonextensive entropic form (1) naturally leads to normalized $q$-expectation values and to escort distributions.

Let us further elaborate on Eq. (7). It can be also rewritten in a more symmetric form, namely as
\begin{equation}
1+(1-q)S_q(\{p_i\}) = [1+(1-q)S_q(\{\pi_k\}][1+(1-q)\langle \langle S_q(\{\ p_i / \pi_k\})  \rangle \rangle_q]
\end{equation}
Since the {\it Renyi entropy} (associated with the probabilities $\{p_i\}$) is defined as $S_q^R(\{p_i\}) \equiv (\ln \sum_{i=1}^W p_i^q)/(1-q) $, we can conveniently define the (monotonically increasing) function ${\cal R}_q[x] \equiv \ln[1+(1-q)x]/[1-q]=\ln \{[1+(1-q)x]^{[1/(1-q)]}\}$ (with ${\cal R}_1[x]=x$), hence, for any distribution of probabilities, we have $S_q^R={\cal R}_q[S_q]$. Eq. (10) can now be rewritten as
\begin{equation}
{\cal R}_q[S_q(\{p_i\})] = {\cal R}_q[S_q(\{\pi_k\})] + {\cal R}_q[\langle \langle S_q(\{\ p_i / \pi_k\})  \rangle \rangle_q]\;,
\end{equation}
or equivalently,
\begin{equation}
S_q^R(\{p_i\}) = S_q^R(\{\pi_k\}) + {\cal R}_q[\langle \langle  {\cal R}_q^{-1}[S_q^R(\{\ p_i / \pi_k\})]  \rangle \rangle_q]\;,
\end{equation}
where the inverse function is defined as ${\cal R}_q^{-1}[y] \equiv [(e^y)^{(1-q)}-1]/[1-q]$ (with ${\cal R}_1^{-1}[y]=y$). Notice that, in general, ${\cal R}_q[\langle \langle ... \rangle \rangle_q] \ne \langle \langle {\cal R}_q[...] \rangle \rangle_q$.

Everything we have said in this Section is valid for {\it arbitrary} partitions (in $U$ nonintersecting subsets) of the ensemble of $W$ possibilities. Let us from now on address the particular case where the $W$ possibilities correspond to the joint possibilities of two subsystems $A$ and $B$, having respectively $W_A$ and $W_B$ possibilities (hence $W=W_A W_B$). Let us denote by $\{p_{ij}\}$ the probabilities associated with the total system $A+B$, with $i=1,\;2,\;...,\;W_A$, and  $j=1,\;2,\;...,\;W_B$. The {\it marginal probabilities} $\{ p_i^A\}$ associated with subsystem $A$ are given by $p_i^A = \sum_{j=1}^{W_B} p_{ij}$, and those associated with subsystem $B$ are given by $p_j^B = \sum_{i=1}^{W_A} p_{ij}$. $A$ and $B$ are said to be {\it independent} if and only if $p_{ij} = p_i^Ap_j^B\;(\forall (i,j))$. We can now identify the $U$ subsets we were previously analyzing with the $W_A$ possibilities of subsystem $A$, hence the probabilities $\{\pi_k \}$ correspond respectively to $\{ p_i^A\}$. Consistently, Eq. (7) implies now
\begin{equation}
S_q[A+B] = S_q[A] + S_q[B|A] + (1-q)S_q[A]S_q[B|A]\;,
\end{equation}
where $S_q[A+B] \equiv S_q(\{p_{ij}\})$, $S_q[A] \equiv S_q(\{p_i^A\})$ and the conditional entropy 
\begin{equation}
S_q[B|A] \equiv \frac{\sum_{i=1}^{W_A} (p_i^A)^q S_q[B|A_i]}{\sum_{i=1}^{W_A} (p_i^A)^q} \equiv \langle \langle S_q[B|A_i \rangle \rangle_q\;,
\end{equation}
where
\begin{equation}
S_q[B|A_i] \equiv \frac{1- \sum_{j=1}^{W_B} (p_{ij}/p_i^A)^q}{q-1}\;\;\;(i=1,\;2,\;...,\;W_A)\;,
\end{equation}
with $\sum_{j=1}^{W_B} (p_{ij}/p_i^A)=1$. Symmetrically, Eq. (13) can be also written as
\begin{equation}
S_q[A+B] = S_q[B] + S_q[A|B] + (1-q)S_q[B]S_q[A|B]\;.
\end{equation}
If $A$ and $B$ are independent, then $p_{ij}=p_i^Ap_j^B\;(\forall )i,j))$, hence $S_q[A|B] = S_q[A]$ and $S_q[B|A] = S_q[B]$, therefore both Eqs. (13) and (16) yield the well known pseudo-extensivity property of the nonextensive entropy $S_q$, namely
\begin{equation}
S_q[A+B] = S_q[A] + S_q[B] + (1-q)S_q[A]S_q[B]\;.
\end{equation}
We thus see that Eqs. (13) and (16) nicely compress into {\it one} property two important properties of the entropic form $S_q$, namely Eq. (2) and Eq.(17). The axiomatic implications of these relations have been discussed by Abe\cite{abeaxiom}. 

\section{QUANTUM CONDITIONAL NONEXTENSIVE ENTROPY, QUANTUM ENTAGLEMENT AND SEPARABILITY}

Let us now address the quantum version of the expressions discussed in the previous Section. Eq. (1) is generalized into
\begin{equation}
S_q=\frac{1- Tr \rho^q}{q-1}\;\;\;(q \in \cal {R}; \; $Tr$ \;\rho=$1$;\;$S$_1=-$Tr$\; \rho \ln \rho)\;, 
\end{equation}
where $\rho$ is the density operator. Eq. (1) is recovered as the particular instance when $\rho$ is diagonal. 

Let us assume now that $\rho$ is the density operator associated with a composed system $A+B$. Then the marginal density operators are given by  $\rho_A \equiv Tr_B\; \rho$ and $\rho_B \equiv Tr_A\; \rho$ ($Tr_A \;\rho_A=Tr_B \;\rho_B=1$). The systems $A$ and $B$ are said to be {\it uncorrelated } (or {\it independent}) if and only if 
\begin{equation}
\rho = \rho_A \otimes \rho_B \;.
\end{equation}
Otherwise they are said to be {\it correlated}. Two correlated systems are said to be {\it separable} (or {\it unentangled}) if and only if it is possible to write $\rho$ as follows:

\begin{equation}
\rho = \sum_{i=1}^W p_i \;\rho_A^{(i)} \otimes \rho_B^{(i)} \;\;\;\; (p_i \ge 0\; \forall i; \;\sum_{i=1}^W p_i=1).
\end{equation}
The limiting case of independency is recovered for {\it certainty}, i.e., if all $\{p_i\}$ vanish excepting one which equals unity. {\it Nonseparability} (or {\it entanglement}) is at the basis of the amazing phenomena mentioned in the Introduction and, as already mentioned, at the center of the admissibility of a {\it local realistic} description of the system in terms of hidden variables. 
Correlation is a concept which is meaningful both classically and quantically. Entanglement, in the present sense, is meaningful only within quantum mechanics. The characterization of quantum entanglement is not necessarily simple to implement, since it might be relatively easy in a specific case to exhibit the form of Eq. (20), but it can be nontrivial to prove that it {\it cannot} be presented in that form. Consequently, along the years appreciable effort has been dedicated to the establishment of general operational criteria, preferentially in the form of necessary and sufficient conditions whenever possible. Peres\cite{peres} pointed out a few years ago a {\it necessary} condition for separability, namely the nonnegativity of the partial transpose of the density matrix. In some simple situations (like the simple mixed state of two spin 1/2  we shall focuse on later on) Peres' criterion is now known to also be a {\it sufficient} condition. But, as soon as the case is slightly more complex (e.g., $3 \times 3$ or $2 \times 4$ matrices) it is known now to be {\it insufficient}.

It is within this scenario that Abe and Rajagopal\cite{aberajagopal} recently proposed a different condition, presumably a {\it necessary} one, based on the nonextensive entropic form $S_q$ given in Eq. (18). Let us summarize the idea. The quantum version of conditional probabilities is uneasy to formulate in spite of being so simple within a classical framework (see Section II). The difficulties come from the fact that generically $\rho$ does not commute with either $\rho_A$ or $\rho_B$. Consistently, the concept of {\it quantum conditional entropy} is a sloppy one. Abe and Rajagopal suggested a manner of shortcutting the difficulty, namely through the adoption, also for quantum systems, of Eqs. (13) and (16)  as they stand. Consequently, the conditional entropies $S_q(A|B)$ and $S_q(B|A)$ are respectively given by
\begin{equation}
S_q(A|B)\equiv \frac{S_q(A+B)-S_q(B)}{1+(1-q)S_q(B)},
\end{equation}
and
\begin{equation}
S_q(B|A)\equiv \frac{S_q(A+B)-S_q(A)}{1+(1-q)S_q(A)},
\end{equation}
where $S_q(A+B) \equiv S_q(\rho)$, $S_q(A) \equiv S_q(\rho_A)$ and $S_q(B) \equiv S_q(\rho_B)$. Obviously, for the case of independence, i.e., when $\rho = \rho_A \otimes \rho_B$, these expressions lead to $S_q(A|B)=S_q(A)$ and $S_q(B|A)=S_q(B)$, known to be true also in quantum mechanics.

Both classical and quantum entropies $S_q(A+B)$, $S_q(A)$ and $S_q(B)$ are always nonnegative. This is not the case of the conditional entropies $S_q(A|B)$ and $S_q(B|A)$, which are always nonnegative classically, but which can be negative quantically (see also \cite{cerfadami}). It is therefore natural to expect that separability implies nonnegativity of the conditional entropies. This is the criterium proposed in \cite{aberajagopal}. More over, since these conditional entropies are monotonically decreasing functions of $q$, the strongest case corresponds to the nonnegativity of both $S_{\infty}(A|B)$ and  $S_{\infty}(B|A)$.  

In what follows, we shall illustrate these general ideas on a paradigmatic two spin 1/2 system.
The simplest basis for describing such a system is $|$$\uparrow>_A|$$\uparrow>_B$, $|$$\uparrow>_A|$$\downarrow>_B$, $|$$\downarrow>_A|$$\uparrow>_B$ and $|$$\downarrow>_A|$$\downarrow>_B$. All these states clearly are unentangled. Another popular basis is the {\it triplet} constituted by $|$$\uparrow>_A|$$\uparrow>_B$, $|$$\downarrow>_A|$$\downarrow>_B$ and $|\Psi^+$$> \equiv \frac{1}{\sqrt 2}(|$$\uparrow>_A|$$\downarrow>_B$$+|$$\downarrow>_A|$$\uparrow>_B)$, together with the {\it singlet} $|\Psi^-$$> \equiv \frac{1}{\sqrt 2}(|$$\uparrow>_A|$$\downarrow>_B$$-|$$\downarrow>_A|$$\uparrow>_B)$. Here, we start having entanglement. Finally we have the Bell basis, convenient for a variety of experimental situations, constituted by $|\Psi^{\pm}$$> \equiv \frac{1}{\sqrt 2}(|$$\uparrow>_A|$$\downarrow>_B$$\pm|$$\downarrow>_A|$$\uparrow>_B)$ and  $|\Phi^{\pm}$$> \equiv \frac{1}{\sqrt 2}(|$$\uparrow>_A|$$\uparrow>_B \pm |$$\downarrow>_A|$$\downarrow>_B)$. Each state of this basis is fully entangled. They satisfy
\begin{equation}
|\Phi^+$$><$$\Phi^+|+|\Phi^-$$><$$\Phi^-|+|\Psi^+$$><$$\Psi^+|+|\Psi^-$$><$$\Psi^-|=\hat1_{A+B} \equiv \hat1_A \otimes \hat1_B 
\end{equation}
with $Tr \; \hat 1_{A+B}=2\; Tr\; \hat 1_A= 2\; Tr \;\hat1_B= 4$.

Let us first assume that our bipartite system is in a simple mixed state (so called Werner-Popescu \cite{werner,popescu} state), namely
\begin{equation}
\rho=\frac{1-x}{4}(|\Phi^+$$><$$\Phi^+|+|\Phi^-$$><$$\Phi^-|+|\Psi^+$$><$$\Psi^+|)+\frac{1+3x}{4}|\Psi^-$$><$$\Psi^-|
\end{equation}
or equivalently
\begin{equation}
\rho=\frac{1-x}{4}\;\hat1_{A+B}+x\;|\Psi^-$$><$$\Psi^-| \;\;\;(0 \le x \le 1),
\end{equation}
where we have used Eq. (23). For $x=1$ and $x=0$ we have the fully entangled EPR state and the fully random one respectively. The question arises: up to what value of $x$ is separability possible? The use of the Bell inequality yields that the threshold cannot exceed $1/\sqrt{2} \simeq 0.71$. The use of the $\alpha$-entropic inequality \cite{horodecki1} yields a more severe restriction, namely that it cannot exceed  $1/\sqrt{3} \simeq 0.58$. The strongest result, i.e., the necessary and sufficient condition, was finally found  (by imposing the nonnegativeness of the partial transpose of the density matrix) by Peres \cite{peres}, and it is $x_c=1/3$. As already mentioned, this Peres' recipe is known to be a {\it necessary} condition for all sytems, and has been shown to be also {\it sufficient} for this $2 \times 2$ system (as well as for some $2 \times 3$ systems), whereas it is known to be {\it insufficient} for $3 \times 3$ and $2 \times 4$ or more complex systems \cite{horod,bruss}. It is this $x_c=1/3$ result that Abe and Rajagopal \cite{aberajagopal} elegantly reproduced. 

Along the same lines, it was recently considered\cite{sethmichel} a more general mixed case, namely
\begin{equation}
\rho_{A+B}=\frac{1-x}{4}\;|\Phi^+$$><$$\Phi^+|
+ \frac{1-y}{4}\;|\Phi^-$$><$$\Phi^-| + \frac{1-z}{4}\;|\Psi^+$$><$$\Psi^+| + \frac{1+x+y+z}{4}\;|\Psi^-$$><$$\Psi^-|
\end{equation}
or equivalently
\begin{equation}
\rho_{A+B}=\frac{1}{4}\;\hat1_{A+B}   -\frac{x}{4}\;|\Phi^+$$><$$\Phi^+|-\frac{y}{4}\;|\Phi^-$$><$$\Phi^-|-\frac{z}{4}\;|\Psi^+$$><$$\Psi^+|
+(x+y+z)\;|\Psi^-$$><$$\Psi^-|\;,
\end{equation}
with $x,y,z \le 1$. Eqs. (24) and (25) are reproduced in the $x=y=z$ case. The pure states $|\Phi^+$$>$, $|\Phi^-$$>$, $|\Psi^+$$>$ and $|\Psi^-$$>$ (EPR state) respectively correspond to $(x,y,z)=(-3,1,1),\;(1,-3,1),\;(1,1,-3)$ and $(1,1,1)$. For this state, it was found\cite{sethmichel}
\begin{equation}
S_q(A+B)=\frac{1}{1-q}\Bigl[\Bigl(\frac{1-x}{4}\Bigr)^q+  \Bigl(\frac{1-y}{4}\Bigr)^q+\Bigl(\frac{1-z}{4}\Bigr)^q+\Bigl(\frac{1+x+y+z}{4}\Bigr)^q-1 \Bigr].
\end{equation}
and
\begin{equation}
S_q(A)=S_q(B)=\frac{2^{1-q}-1}{1-q}\;.
\end{equation}
Substituting these two last expressions into Eqs. (21) and (22) we obtain $S_q(A|B)=S_q(B|A)$ as an explicit function of $(x,\;y,\;z;\;q)$ (See Figs. 1 and 2). Both 
$S_q(A+B)$ and $S_q(A|B)=S_q(B|A)$ are invariant under the transformations $(x,y,z) \rightarrow (x,z,y)$, $(x,y,z) \rightarrow (-x-y-z,x,y)$, and the analogous ones. $S_q(A|B)=S_q(B|A)=0$ implies
\begin{equation}
\Bigl(\frac{1-x}{4}\Bigr)^q+  \Bigl(\frac{1-y}{4}\Bigr)^q+\Bigl(\frac{1-z}{4}\Bigr)^q+\Bigl(\frac{1+x+y+z}{4}\Bigr)^q = \frac{1}{2^{q-1}}\;.
\end{equation}
In the limit $q \rightarrow \infty$, this relation implies
\begin{equation}
x+y+z=1\;.
\end{equation}
The same condition can be straightforwardly found through Peres' partial transpose procedure. It was advanced in \cite{sethmichel} that perhaps the conditional entropic criterium (i.e., nonnegativity of both $S_{\infty}(A|B)$ and  $S_{\infty}(B|A)$) was a {\it sufficient} condition, in addition of being a {\it necessary} one. The analysis\cite{alcaraz} of further and more complex systems (e.g., the full $2 \times 2$ case, whose associated density matrix involves 15 independent real numbers, as well as selected many-spin mixed states) has shown counterexamples that in general preclude such possibility. In all cases that we have fully worked out, the entropic criterium has either provided the same result as the Peres' one, or has proved to be slightly less restrictive. Although we have no proof that it cannot be the other way around (i.e., Peres' criterium being less restrictive than the entropic one), we have not encountered such an example (see \cite{horodeckitsallis} for more details). In some cases of increasingly many spins, both criteria have produced the same result, presumably the asymptotically exact answer. Details will be published elsewhere\cite{alcaraz}. As a general trend, it seems that the transposed matrix criterium {\it overestimates the separable region} (i.e., {\it underestimates the quantum entangled region}) less or equal than the conditional entropy criterium. It could well be that whenever both criteria produce the same result, this result is the exact answer.

Let us now focus on the critical-phenomenon-like scenario we have previously announced. Its possibility was introduced in \cite{sethmichel} but it was not exhibited for any case. Let us do this here for the $x=y=z$ particular case. 
As a function of $x$ we have numerically calculated the value $q_0$ at which $S_q(A|B)$ vanishes and its associated derivative $\partial S_q(A|B) / \partial q|_{q=q_0}$, as well as the
inflexion point of $S_q(A|B)$ versus $q$ (see examples in Figs. 2 and 3). The abcissa and ordinate corresponding to the inflexion point are noted respectively $q_I$ and $S_{q_I}$. These quantities, as well as $q_0$ and $\partial S_q(A|B) / \partial q|_{q=q_0}$, are represented in Fig. 4. 

We see that $q_0$ diverges for $0 \le x \le 1/3$ ({\it separable state}), and decreases from infinity to zero when $x$ increases from 1/3 to unity ({\it entangled state}). It is therefore convenient to define $\mu \equiv 1/(1+q_0)$, which increases from zero to unity when $x$ increases from 1/3 to unity (see Fig. 4(a)). This line plays a role analogous to a critical line, separating the unit square in the $(x,\mu)$ space into two regions, respectively corresponding to the separable and entangled states. This possible interpretation is one of the two scenarios advanced in \cite{sethmichel}, and seems more appropriate than the other scenario, in which a line in the $(x,\mu)$ space was thought as a critical phenomenon order parameter. In the present scenario, we need to define an order parameter $M(x,\mu)$, as well as a ``thermodynamically conjugate" parameter $H$, such that the ``susceptibility" $\chi \equiv \partial M(x,\mu,H)/\partial H|_{H=0}$ diverges (presumably with the same critical exponent) when the critical line is approached from {\it both} sides. Furthermore, we would need to identify the ``symmetry" which is broken at the critical line. Naturally, $M(x,\mu,0)$ is expected to vanish in the entire separable state, and increase from zero at the critical line (in the $(x,\mu)$ space) up to unity for $(x,\mu)=(1,0)$. It is even plausible that $M(1,\mu,0)=1\;(\forall \mu)$, whereas $M(x,0,0)$ would continuously increase from zero to unity when $x$ increases from 1/3 to unity. If this is so, we would have a nonuniform convergence feature, more precisely $\lim_{\mu \rightarrow 1} \lim_{x \rightarrow 1} M(x,\mu,0)= 1  $ while $\lim_{x \rightarrow 1} \lim_{\mu \rightarrow 1} M(x,\mu,0)= 0  $. At the other end of the critical line, convergence would be uniform, i.e., $\lim_{\mu \rightarrow 0} \lim_{x \rightarrow 1/3} M(x,\mu,0)= \lim_{x \rightarrow 1/3} \lim_{\mu \rightarrow 0} M(x,\mu,0)= 0$. Finally, quantum entanglement should disappear in all cases, i.e., at all values of $(x,\mu)$, in the classical limit, i.e., when Planck constant $h$ vanishes. Consistently, it should be possible to present the quantity $M(x,\mu,H)$ as the proportionality factor of some other intrinsically quantum quantity ${\cal M}(x,\mu,H,h)$, i.e., we should have ${\cal M}(x,\mu,H,h)=h M(x,\mu,H)$. Indeed, such a form would imply that ${\cal M}(x,\mu,H,h)$ vanishes for $h =0$, and is finite for {\it any} value of $h \ne 0$. Obviously, this scenario needs further analysis, which would be welcome.

\section{CONCLUSIONS}

We have analyzed classical systems and have seen, along Khinchin-Abe lines\cite{abeaxiom}, how the $q$-generalization of Shannon's celebrated property (Eq. (2)) naturally leads to a pseudo-extensive property involving conditional entropies (Eqs. (13) and (16)), which in turn recovers as particular case the well known pseudo-extensive property (Eq. (17)) of entropies associated with independent systems. Such conditional entropies, when maintained in form for the quantum case\cite{aberajagopal}, enable what appears to be a necessary condition for separability. This condition seems to generically be coincident or slightly less restrictive than Peres' criterium based on the partially transposed density matrix. The latter, also a necessary condition in general, occasionally is sufficient as well, like in the simple mixed state of two spins 1/2 (see \cite{GHZ,horodecki2,peres,horod,bruss} for details). 

For the two spins 1/2 mixed state characterized by $x=y=z$, both entropic and transposed matrix criteria yield the exact answer of the problem. 
We have exhibited that, through special values of the conditional entropy $S_q(A|B)=S_q(B|A)$ as a function of $q$, quantum separability (hence local realism) presents some analogies with standard critical phenomena. Indeed, a line quite analogous to a critical one emerges in the $(x,q)$ space (with $ 0 \le x \le 1$), which becomes divided into two regions, namely the separable one and the entangled one (which includes as its most entangled state the EPR one). Within this standpoint, it becomes clear that the Boltzmann-Gibbs-Shannon entropy ($q=1$) is a concept too poor for properly discussing quantum entanglement, a conclusion recently reached also by Brukner and Zeilinger\cite{others} from a different path. 
Obviously, in the more complex cases where the present entropic criterium does not provide the exact answer (but only an upper bound of it), the present considerations should have to be appropriately adapted. But, on qualitative grounds, the situation is expected to be essentially the same.

An appreciable amount of arguments are now available in the literature which connect quantum entanglement and thermodynamics \cite{others,zurek2,horodecki1,horodecki2,popescu2,horodecki3,horodecki4}. 

The possibility explored here (alternative to the one preliminary sketched in \cite{sethmichel}) where curves in the $(q,x)$ space appear as analogous to critical lines emerged during discussions with  R. Horodecki, P. Horodecki and M. Horodecki, to whom one of us (CT) is grateful. Enlightening remarks from F. C. Alcaraz, E.M.F. Curado and E.P. Borges are acknowledged as well.

\centerline{{\bf FIGURE CAPTIONS}}

\bigskip
\noindent
Fig. 1.~
The conditional entropy $S_q(A|B)=S_q(B|A)$ versus $(x,y,z)$ for typical values of $q$: (a) $q=1/2$ for the solid lines, $q=2$ for the dashed lines, and $q=5$ for the dotted lines, along the directions $(x,0,0)$, $(x,x,0)$ and $(x,x,x)$ from top to bottom; (b) For $(x,y,z)$ along the edge joining $|\Phi^+$$>$ and $|\Psi^-$$>$ or, equivalently, $|\Phi^+$$>$ and $|\Phi^-$$>$ (notice the symmetry with regard to the $x=-1$ axis). From \cite{sethmichel}. 

\bigskip
\noindent
Fig. 2.~
The conditional entropy $S_q(A|B)=S_q(B|A)$ versus $q$, for typical values of $(x,y,z)$. 
The curve which for $q>0$ is the uppermost is given by $(2^{1-q}-1) /(1-q)$. The lowest curve is given by $-(2^{q-1}-1)/(q-1)$. Notice that six interesting nonuniform convergences occur at $q=0$, namely when (i) the $(x,0,0)$ curves approach, for $x \rightarrow 1$, the $(1,0,0)$ curve; (ii)  the $(x,x,0)$ curves approach, for $x \rightarrow 1$, the $(1,1,0)$ curve; (iii) the $(x,x,x)$ curves approach, for $x \rightarrow 1$, the $(1,1,1)$ curve; (iv) the $(1,x,0)$ curves approach, for $x \rightarrow 1$, the $(1,1,0)$ curve; (v) the $(1,x,x)$ curves approach, for $x \rightarrow 1$, the $(1,1,1)$ curve; (vi) the $(1,1,x)$ curves approach, for $x \rightarrow 1$, the $(1,1,1)$ curve. For $q<0$, all curves, excepting the $(1,1,1)$ one, have positive values and curvatures. The $(1,1,1)$ curve is everywhere negative both in value and curvature. From \cite{sethmichel}. 

\bigskip
\noindent
Fig. 3.~
The conditional entropy $S_q(A|B)=S_q(B|A)$ versus $q$, for typical values of $x$ (from top to bottom $x=0,\;1/3,\;...,\;0.9$). The $x$-dependent inflexion points (full cercles) are located at $(q_I,S_{q_I})$.  

\bigskip
\noindent
Fig. 4.~
From the curves indicated in Fig. 3 we obtain the following $x$-dependences of: (a) $\mu \equiv 1/(1+q_0)$ (in the limit $x \rightarrow 1/3 + 0$, we obtain  $q_0 \sim \frac {\ln 4}{3x - 1}$, and, in the limit $x \rightarrow 1$, we obtain $q_0 \sim \frac{\ln 3}{\ln[2/(1-x)]}$); (b) $D \equiv \partial S_q(A|B) / \partial q|_{q=q_0}$ (in the limit $x \rightarrow 1$ we obtain $D \sim \frac{1}{2} \ln \frac{1-x}{2}$); (c) $1/q_I$ (in the limit $x \rightarrow 1/3 + 0$, we obtain $1/q_I \sim a(x-1/3)$ with $a \simeq 1.15$); (d) $S_{q_I}$. The line shown in (a) can be thought as playing the role of a critical line, the {\it separable} ({\it entangled}) phase being the one in the neigborhood of the $(x,\mu)=(0,1)$ ($(x,\mu)=(1,0)$) point. The Peres' critical value $x=1/3$ corresponds to the $q_0 \rightarrow \infty$ limit of this line (which also corresponds to the $q_I \rightarrow \infty$ limit).

\vfill

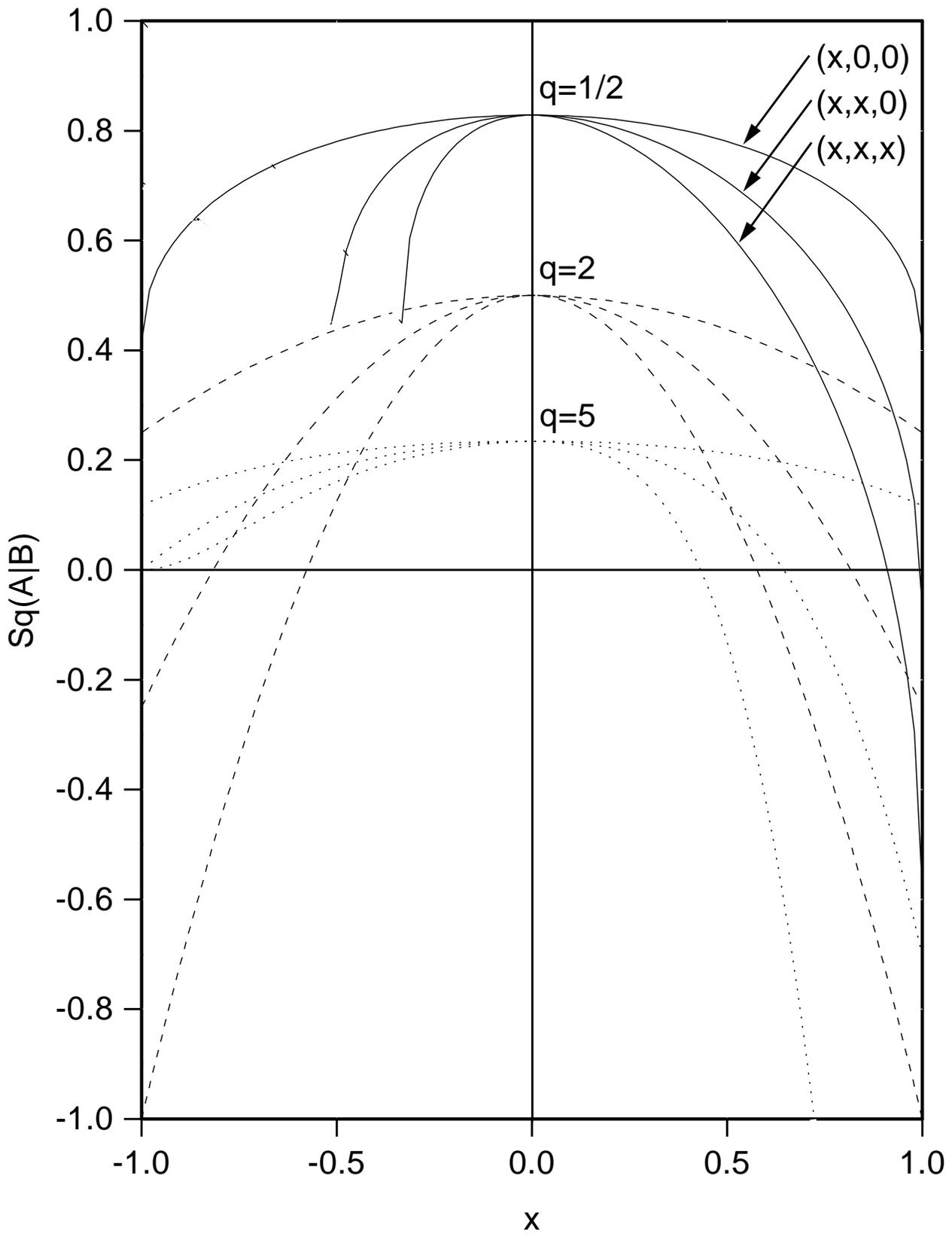
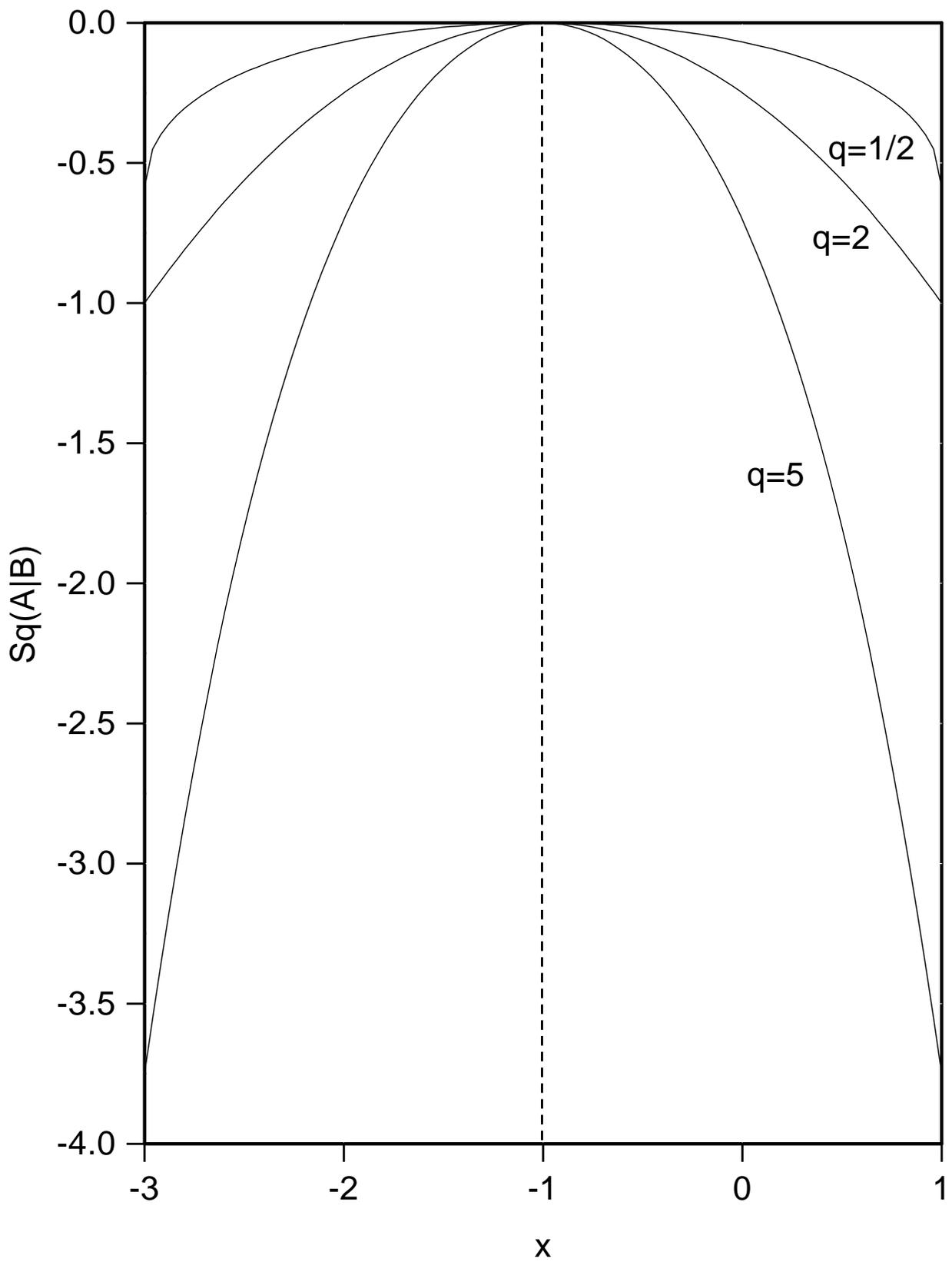
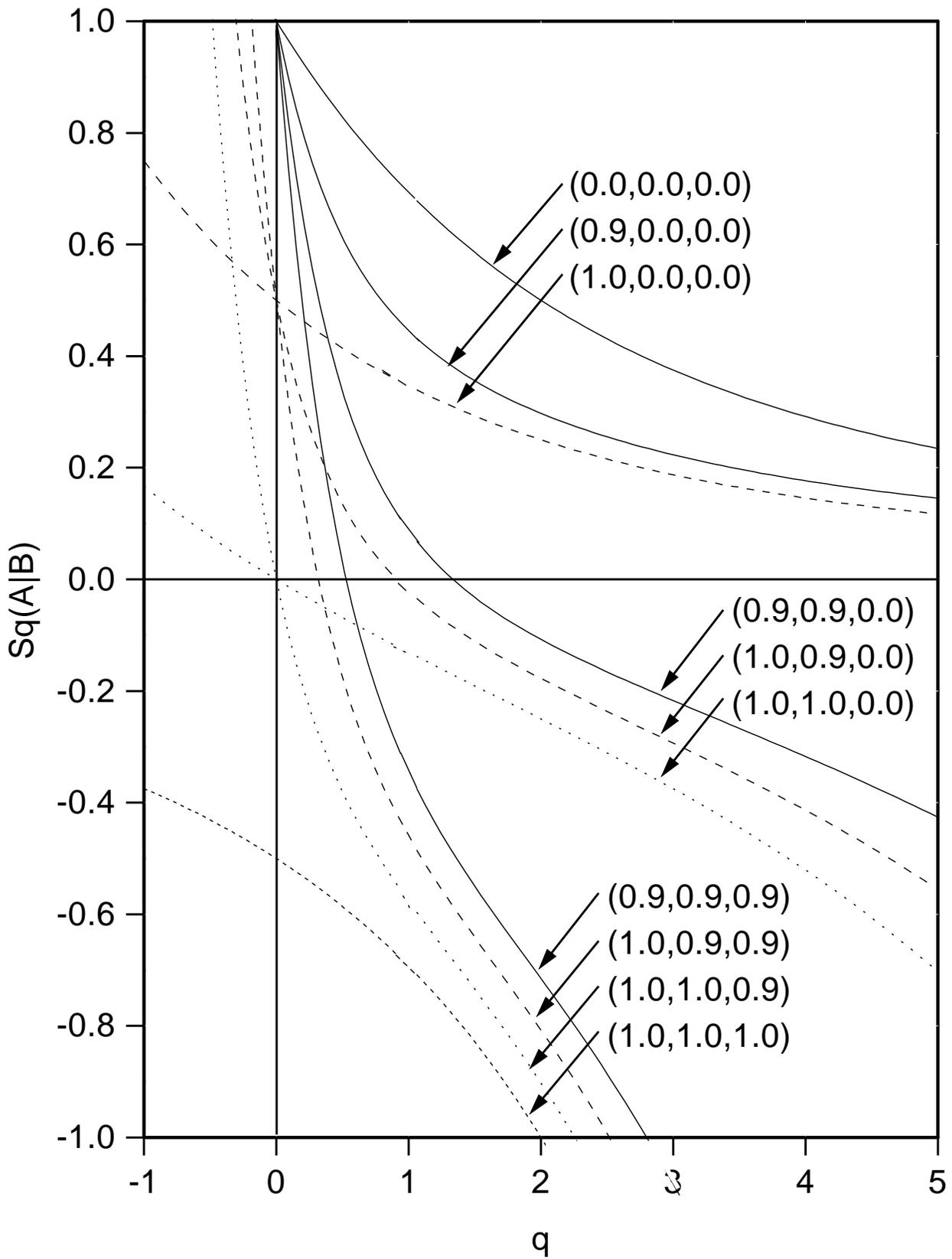
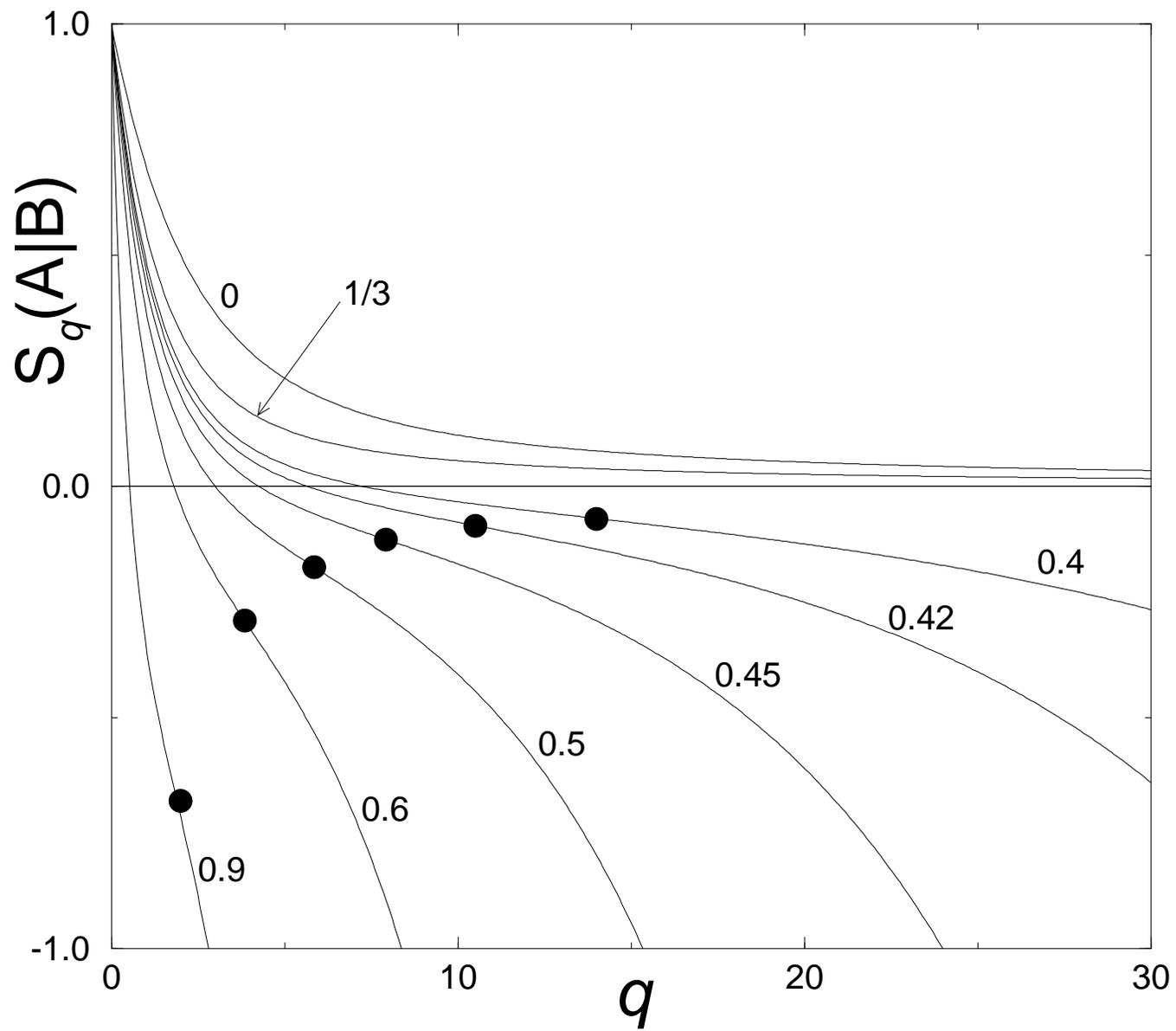
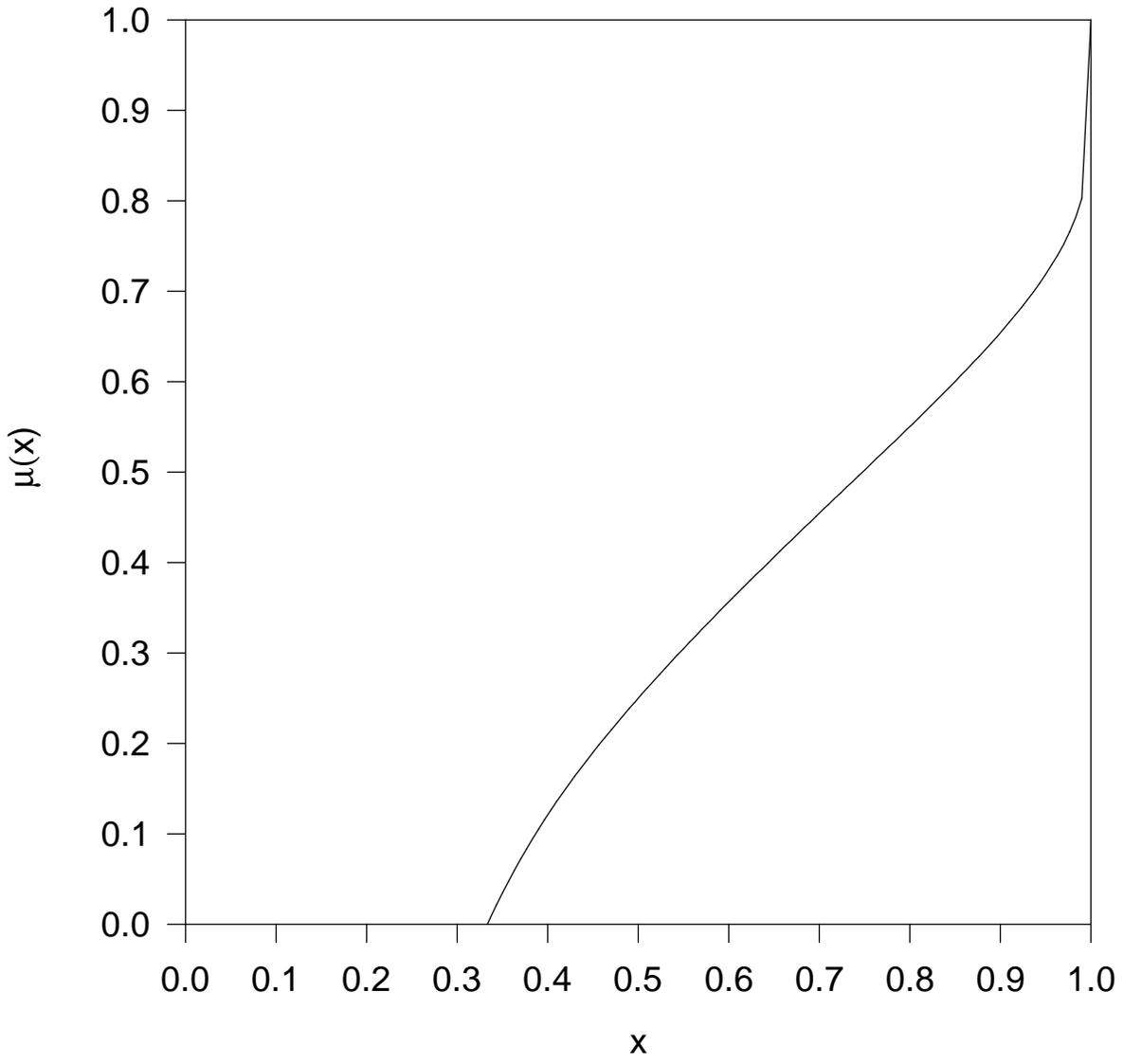
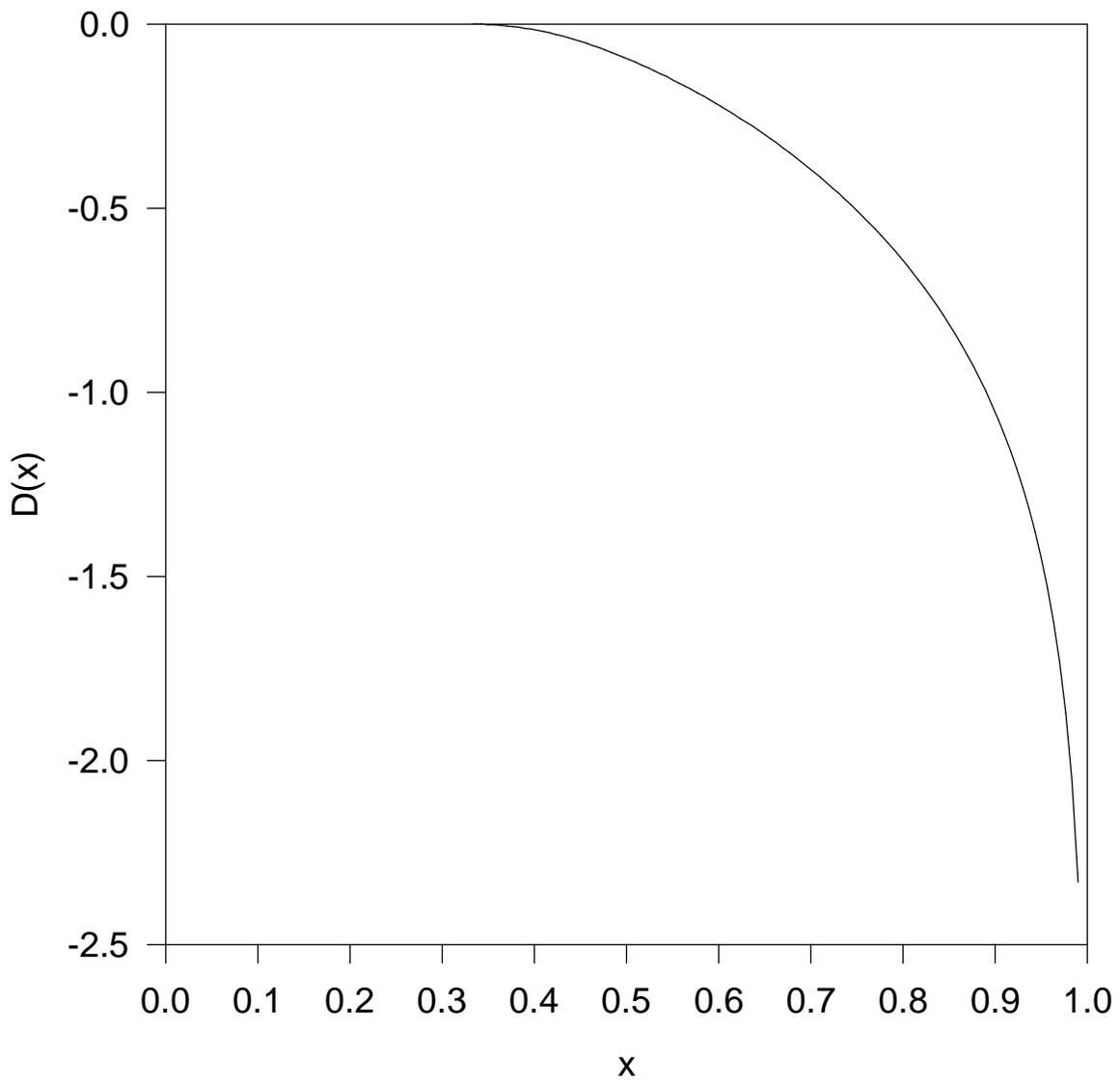
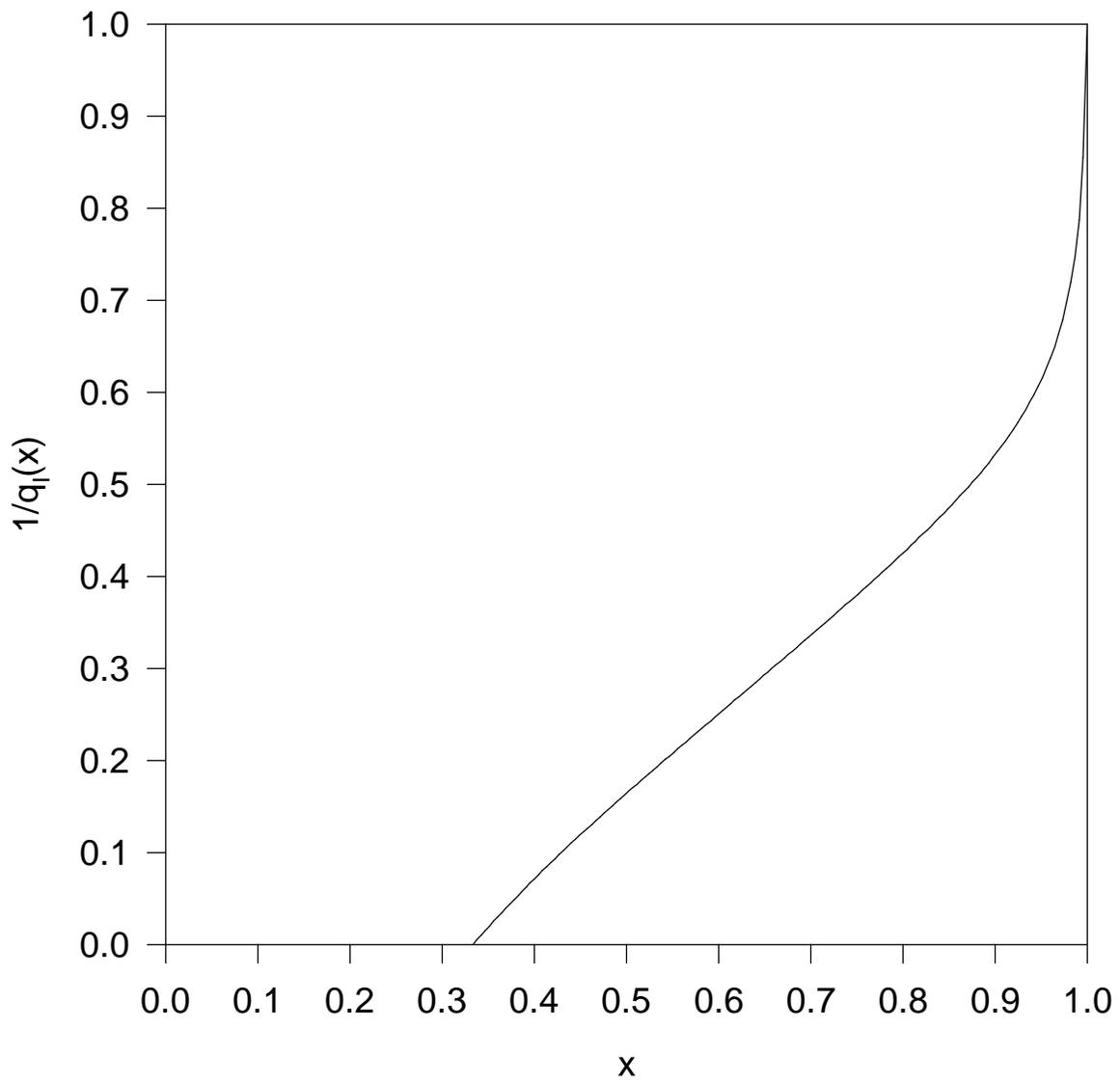
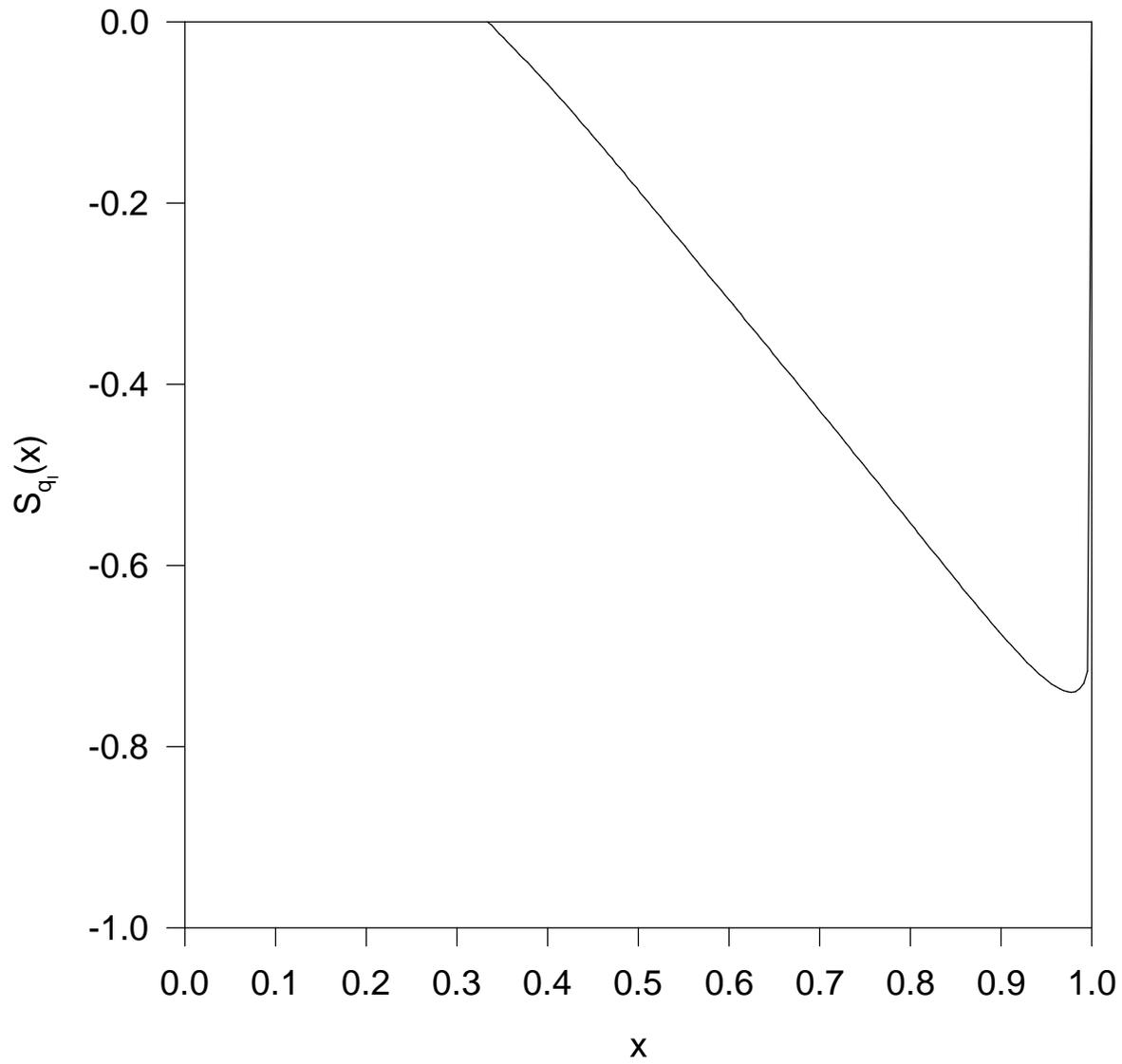
\end{document}